\documentclass[11pt,oneside]{article}
\usepackage{academicons, lastpage, booktabs}
\usepackage[dvipsnames]{xcolor}
\usepackage[english]{babel}
\usepackage{amsfonts,amssymb,amsmath,amsthm}
\usepackage[absolute,overlay]{textpos}
\usepackage{float}
\usepackage{fancyhdr, graphicx}
\usepackage{enumerate}
\usepackage[margin=1in]{geometry}
\usepackage{tikz}
\usepackage{layout}
\usepackage{titlesec}
\usepackage{titling}
\usepackage{authblk}
\usepackage{xcolor, colortbl}
\usepackage{geometry}
\usepackage{siunitx} 
\usepackage{hyperref}
\usepackage{eso-pic}
\usepackage{marginnote}
\usepackage{doi}

\definecolor{coolblack}{rgb}{0.0, 0.18, 0.30}
\definecolor{carmine}{rgb}{0.59, 0.0, 0.09}
\definecolor{myblue}{rgb}{0.0, 0.48, 0.65}
\definecolor{cadetgrey}{rgb}{0.57, 0.64, 0.69}
\definecolor{lightgray}{gray}{0.9}

\hypersetup{
    colorlinks=true,
    linkcolor=coolblack,
   filecolor=magenta,      
    urlcolor=cyan,
    citecolor=coolblack,
    pdfhighlight=/P,
                 }
                                  
\usepackage{libertinus}
\usepackage[T1]{fontenc}

\usepackage{orcidlink}

\usepackage{caption}
\usepackage{fancyhdr}

\titleformat*{\section}{\Large\bfseries\color{black}}
\titleformat*{\subsection}{\large\bfseries\color{black}}
\titleformat*{\subsubsection}{\bfseries\color{black}}

\makeatletter
\renewcommand\@makefntext[1]{\leftskip=0em\hskip0em\@makefnmark#1}
\makeatother
\newcommand\blfootnote[1]{
  \begingroup
  \renewcommand\thefootnote{}\footnote{#1}%
  \addtocounter{footnote}{-1}
  \endgroup
}

\providecommand{\keywords}[1]
{
  \small	
  \textbf{\textsf{Keywords---}} #1
}

\renewcommand{\tableautorefname}

\newcommand{\mycomment}[1]{} 
\newcommand*\mytitle{Impurity Parallel Velocity Gradient instability} 
\newcommand*\myauthor{Bourgeois \textit{et al.}} 
\newcommand*\mycorrmail{maxime.lesur@univ-lorraine.fr} 
\newcommand*\myDOI{10.46298/ops.13628} 
\newcommand*\myarxiv{2405.08472} 
 
\newcommand*\myvolume{1} 
\newcommand*\myyear{2024} 
\newcommand*\mynumber{1 } 
\newcommand*\mydaterec{May 20, 2024}
\newcommand*\mydaterevised{August 16, 2024}
\newcommand*\mydateaccepted{October 05, 2024}
\newcommand*\mydatepublished{\today}

\newlength{\myshift}
\def\vec#1{{\boldsymbol{#1}}}

\def\vb#1{{\mbox{\boldmath$#1$}}}
\def\pd#1#2{\frac{\partial #1}{\partial #2}}

\def\bdot{\,\vb{\cdot}\,}

\def\grad{{\boldsymbol{\nabla}}}

\def\dotp#1#2{{\vec{#1}\bdot\vec{#2}}}
\def\crop#1#2{{\vec{#1}\times\vec{#2}}}

\def\d{{\mathrm{d}}}

\newcommand{\VE}{\vec{u_E}}
\newcommand{\Vps}{\vec{u_{p,s}}}

\newcommand{\GradV}{\overline{ \nabla u_{\parallel} }}

\newcommand{\ex}{\vec{\hat{e_x}}}
\newcommand{\ey}{\vec{\hat{e_y}}}
\newcommand{\ez}{\vec{\hat{e_z}}}

\usepackage[firstpage=true]{background}
\backgroundsetup{contents={\includegraphics[width=1.\paperwidth]{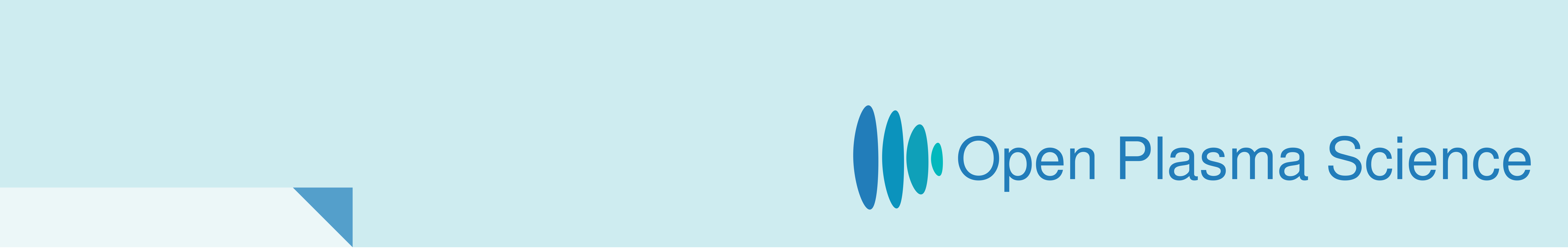}},scale=1,placement=top,opacity=1}

\pretitle{\vspace{-00pt}      \begin{flushleft}   \Large  \color{black} \selectfont  } 
     
\title{\textbf{\mytitle}}

\posttitle{\par \end{flushleft} } 

\preauthor{\vspace{-00pt} \begin{flushleft}  \color{black} \selectfont} 

\author[1,2]{Jeanne Bourgeois}
\author[,3,4]{Maxime Lesur\orcidlink{0000-0001-9747-5616}$^{*}$}
\author[3]{Guillermo Cuerva Lazaro}
\author[5,6]{Yusuke Kosuga\orcidlink{0000-0002-4075-5542}}

\affil[1]{\small Ecole Polytechnique, Institut Polytechnique de Paris, 91128 Palaiseau, France}
\affil[2]{\small Interdisciplinary Graduate School of Engineering Sciences, Kyushu University, Fukuoka, Japan}
\affil[3]{\small Universit{\'e} de Lorraine, CNRS, Institut Jean Lamour, UMR 7198, F-54000 Nancy, France}
\affil[4]{\small Institut Universitaire de France (IUF), Paris, France}
\affil[5]{\small Research Institute for Applied Mechanics, Kyushu University, Kasuga, Fukuoka 816-8580, Japan}
\affil[6]{\small Research Center for Plasma Turbulence, Kyushu University, Kasuga, Fukuoka 816-8580, Japan}

\postauthor{\par\end{flushleft}} 

\date{}

\begin{document}

\maketitle
\thispagestyle{empty}
\blfootnote{$^*$ Corresponding author: \textsf{\href{\mycorrmail}{\mycorrmail}}}
\blfootnote{Cite as: \myauthor, \mytitle, \textit{Open Plasma Science}  \myvolume , \mynumber (\myyear), doi: \myDOI}


\reversemarginpar

\marginnote{\begin{flushleft}
\sf \bf \footnotesize \color{coolblack} History
\end{flushleft}}[-\myshift]
\addtolength{\myshift}{-0.5cm} 
\marginnote{
\begin{flushleft} \footnotesize
Received \mydaterec
\end{flushleft}}[-\myshift]

\addtolength{\myshift}{-0.5cm} 
\marginnote{
\begin{flushleft} \footnotesize
Revised \mydaterevised
\end{flushleft}}[-\myshift]

\addtolength{\myshift}{-0.5cm} 
\marginnote{
\begin{flushleft} \footnotesize
Accepted \mydateaccepted
\end{flushleft}}[-\myshift]

\addtolength{\myshift}{-0.5cm} 
\marginnote{
\begin{flushleft} \footnotesize
Published \mydatepublished
\end{flushleft}}[-\myshift]

\addtolength{\myshift}{-1cm} 

\marginnote{\begin{flushleft}
\sf \bf \footnotesize \color{coolblack} Identifiers
\end{flushleft}}[-\myshift]
\addtolength{\myshift}{-0.5cm} 
\marginnote{
\begin{flushleft} \footnotesize
DOI \href{https://doi.org/\myDOI}{\myDOI}
\end{flushleft}}[-\myshift]
\addtolength{\myshift}{-0.5cm} 
\marginnote{
\begin{flushleft} \footnotesize
HAL -
\end{flushleft}}[-\myshift]
\addtolength{\myshift}{-0.5cm} 
\marginnote{
\begin{flushleft} \footnotesize
ArXiv \href{https://arxiv.org/\myarxiv}{\myarxiv} 
\end{flushleft}}[-\myshift]

\addtolength{\myshift}{-1cm} 
\marginnote{\begin{flushleft}
\sf \bf \footnotesize \color{coolblack} Supplementary Material
\end{flushleft}}[-\myshift]
\addtolength{\myshift}{-0.5cm} 
\marginnote{
\begin{flushleft} \footnotesize
-  
\end{flushleft}}[-\myshift]

\addtolength{\myshift}{-1cm} 
\marginnote{\begin{flushleft}
\sf \bf \footnotesize \color{coolblack} Licence
\end{flushleft}}[-\myshift]
\addtolength{\myshift}{-0.5cm} 
\marginnote{
\begin{flushleft} \footnotesize
\href{https://creativecommons.org/licenses/by/4.0/
}{CC BY} 
\end{flushleft}}[-\myshift]
\addtolength{\myshift}{-0.5cm} 
\marginnote{
\begin{flushleft} \footnotesize
\copyright  The Authors
\end{flushleft}}[-\myshift]
\addtolength{\myshift}{-0.5cm}

\marginnote{\begin{flushleft} \includegraphics[width=0.55\marginparwidth]{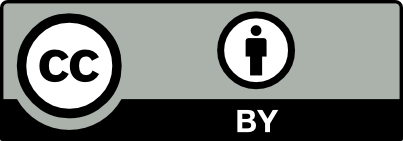}\end{flushleft}}[-\myshift]

\addtolength{\myshift}{11.75cm}
\marginnote{\begin{flushleft}
\sf \footnotesize \color{coolblack} \textsc{\textbf{Vol. \myvolume, No. \mynumber (\myyear)}}
\end{flushleft}}[-\myshift]


\addtolength{\myshift}{-1.5cm} 
  
\marginnote{\begin{flushleft}
\sf \normalsize \color{coolblack} \textsc{\textbf{Regular article}}
\end{flushleft}}[-\myshift]

\begin{flushleft}
\textbf{\textsf{Abstract}}
\vspace{5pt} 
\end{flushleft}

In magnetized plasmas, a radial gradient of parallel velocity, where parallel refers to the direction of magnetic field, can destabilise an electrostatic mode called Parallel Velocity Gradient (PVG). The theory of PVG has been mainly developed assuming a single species of ions. Here, the role of impurities is investigated based on a linear, local analysis, in a homogeneous, constant magnetic field.
To further simplify the analysis, the plasma is assumed to contain only two ion species --- main ions and one impurity species --- while our methodology can be straightforwardly extended to more species. In the cold-ion limit, retaining polarization drift for both main ions and impurity ions, and assuming Boltzmann electrons, the system is described by 4 fluid equations closed by quasi-neutrality. The linearized equations can be reduced to 2 coupled equations: one for the electric potential, and one for the effective parallel velocity fluctuations, which is a linear combination of main ion and impurity parallel velocity fluctuations. This reduced system can be understood as a generalisation of the Hasegawa-Mima model. With finite radial gradient of impurity parallel flow, the linear dispersion relation then describes a new instability: the impurity-modified PVG (i-PVG). 
Instability condition is described in terms of either the main ion flow shear, or equivalently, an effective flow shear, which combines main ion and impurity flow shears. Impurities can have a stabilising or destabilising role, depending on the parameters, and in particular the direction of main flow shear against impurity flow shear. Assuming a reasonable value of perpendicular wavenumber, the maximum growth rate is estimated, depending on impurity mass, charge, and concentration.


\vspace{20pt} 


\begin{flushleft}


\end{flushleft}

\newpage

\newgeometry{ left=20mm, right=20mm,  bottom=3.5cm }

\pagestyle{fancy}
\setlength{\headheight}{15pt}
\fancyhead{} 
\fancyhead[L]{\footnotesize{\href{https://doi.org/\myDOI}{\myDOI}}}
\fancyhead[R]{\footnotesize{\myauthor}}

\fancyfoot{} 
\fancyfoot[R]{\color{coolblack}{ \thepage \hspace{1pt} | \pageref{LastPage}}}
\fancyfoot[C]{\footnotesize{\color{black} Open Plasma Science \textbf{\myvolume}, No.\mynumber (\myyear)}}

\renewcommand{\headrulewidth}{0.4pt}
\renewcommand{\footrulewidth}{0.4pt}
\renewcommand{\footruleskip}{5pt}
\renewcommand\footrule{\hrule width\textwidth}
\renewcommand\headrule{\hrule width\textwidth}


\tableofcontents


\section{Introduction\label{sec:intro}}

In magnetized plasmas, the Parallel Velocity Gradient (PVG), or Parallel Velocity Shear (PVS) is a type of Kelvin-Helmholtz fluid instability driven by a radial gradient of parallel (to magnetic field) plasma flow.
It is sometimes referred to as D’Angelo instability, since D’Angelo investigated a simplified version (simplified by a radial WKB approximation) for a low-temperature plasma in a uniform magnetic field \cite{dangelo65}. The D'Angelo instability was then observed in basic plasma experiments \cite{dangelo66,kaneko03}.

In toroidal magnetic confinement fusion devices, magnetic shear is stabilising \cite{catto73}. However, theory indicates that PVG-driven turbulence may be found in the vicinity of transport barriers \cite{mccarthy02,garbet02}, in the SOL \cite{drake92}, in the presence of strong parallel beam injection, and more readily in spherical tokamaks \cite{chapman12,wang15}.
Even in cases where PVGs are weak, they can have important impact by coupling with Ion-Temperature-Gradient (ITG) turbulence. Theory predicted subcritical ITG-PVG turbulence \cite{newton10,barnes11,schekochihin12,highcock12}, which is consistent with measurements (by Beam Emission Spectroscopy) on MAST \cite{field12}. Strong experimental hints, from fluctuation level and isotropy of correlation length, indicate a significant contribution of PVG-driven turbulence in the edge of the CT-6B tokamak plasma \cite{wang98}. The universality of PVGs in tokamaks remains an open issue.

Parallel flows are also expected to play major roles in linear magnetized plasma devices, such as PANTA (Plasma Assembly for Nonlinear Turbulence Analysis, formerly LMD), where an uphill, near-axis, axial particle flux \cite{kobayashi16} has been measured to be consistent with PVG/drift-waves coupling \cite{inagaki16}, with a regime transition in quantitative agreement with the theoretical linear instability threshold \cite{kosuga15}.

Magnetic confinement fusion plasmas are often contaminated by ion species other than hydrogen isotopes, which are then called as impurities. These can include helium, nitrogen, neon, argon, beryllium, carbon, and tungsten.
Linear magnetized plasma experiments can also include impurities. In particular, a new linear device, called SPEKTRE (Sheath, Plasma Edge \& Kinetic Turbulence Radiofrequency Experiment) \cite{brochard2023spektre} and currently under construction, is partly designed to inject various impurities in a controlled manner and investigate their impact and their dynamics. \\

Our objective here is to investigate the instability driven by a radial shear of the parallel flows of multiple species, which we refer to as impurity-modified PVG (i-PVG). We consider experimental conditions that are relevant to linear plasma experiments with cylindrical geometry such as PANTA and SPEKTRE, and are insightful approximations of tokamak's edge plasma, where the PVG instability is of its most significance. Our local, linearized model can be understood as a modified Hasegawa-Mima model including equilibrium and perturbed parallel velocities of both main ions and impurities. 

In the case of slab geometry with weak magnetic shear, Guo obtained from gyrokinetic equations a linear dispersion relation for PVG with impurities, showing the stabilization of the PVG instability by the presence of impurities and by the increase of electron density gradient \cite{guo19}. Here we adopt a fluid approach, and investigate how the i-PVG threshold, frequency and growth rate depend on parameters such as impurity charge, mass, concentration and parallel flow shear. We further assume a cold-ion limit to provide more clear analysis of the i-PVG. Our fluid approach is supported by this cold ion assumption, plus by the fact that our study focuses on the linear growth of the instabilities and ignores the damped states, so that kinetic effects such as Landau damping can be neglected. Compared to Guo's previous work, our approach provides a more flexible analysis in terms of parameter scan for independent ion and impurity flow shears, and demonstrates that the presence of impurity can be either stabilizing or destabilizing for PVG. The relationship between linear properties and plasma parameters turns out to be somewhat complex, with several non-monotonous dependencies.

\section{Model\label{sec:model}}

The model is based on a local approximation in a Cartesian basis ($\ex$, $\ey$, $\ez$). The magnetic field is assumed homogeneous and constant, $\vec{B}=B \ez$. Hereafter, the parallel and perpendicular subscripts ($\parallel$ and $\perp$) refer to the direction of the magnetic field $\ez$ and the perpendicular plane ($\ex$, $\ey$).
For each species $s$, both equilibrium density gradient and equilibrium parallel velocity gradient are assumed constant and in the $x$ direction: 
\begin{equation}
	\grad{n_{s,0}} \;=\; \frac{n_{s,0}}{L_{n,s}}\ex
\end{equation}
\begin{equation}
	\grad{u_{\parallel,s,0}} \;=\; \frac{u_{\parallel,s,0}}{L_{u,s}}\ex
\end{equation}
where $L_{n,s}$ and $L_{u,s}$ are the density gradient length and parallel velocity gradient length. 

We assume for simplicity that the plasma contains only electrons ($s=e$) and two ion species (one main ion species $s=i$, and one impurity species $s=z$), although the model can be straightforwardly generalized to more species. For each species, mass is noted $m_s$, and charge is noted $q_s=Z_s e$, where $e$ is the elementary charge ($Z_e=-1$ and, for simplicity, we assume that the main ions satisfy $Z_i=1$).

We adopt the cold ion limit, setting for the wave number an approximate maximum limit $\sqrt{T_e/T_s}/\rho_s$, with $s=i$ and $z$, and $\rho_{c,s}$ being the Larmor radius of species $s$. The perpendicular motion of ions (both main ions and impurity ions) is dominated by $E \times B$ drift,
\begin{equation}
	\VE \;=\; \frac{\crop{E}{B}}{B^2}
\end{equation}
and polarization drift,
\begin{equation}
	\vec{u_{p,s}} 
	\;=\; \frac{1}{\omega _{c,s}\, B} \crop{B}{}\frac{\d \VE}{\d t} 
	\;=\; \frac{1}{\omega _{c,s}\, B} \frac{\d \vec{E_\perp}}{\d t}
\end{equation}
This is valid up to first order in terms of the ratio between wave frequency and cyclotron frequency, $\omega / \omega _{c,s}$. Here, as verified a posteriori, $\omega / \omega _{c,i}$ remains as small as a few percent. However, since $\omega_{c,z} / \omega _{c,i}=Z_z m_i/m_z$, the model requires heavy impurities to be sufficiently ionized (unless these impurities are in high enough concentration to bring $\omega$ down to small enough values).

These assumptions yield a system of four fluid equations. Namely, for each of $s=i$ and $z$, the density equation writes as
\begin{equation}
	\label{eq:PVG_cont_imp}
	\pd{n_s}{t} \,+\,  \dotp{\grad{_\perp}} [n_s (\VE + \Vps)] \,+\,  \nabla_\parallel (n_s u_{\parallel,s}) \;=\; 0
\end{equation}
and the parallel flow equation writes as
\begin{equation}
	\label{eq:PVG_mot_imp}
	\pd{u_{\parallel,s}}{t}  \,+\, \dotp{(\VE + \Vps)}{\grad{_\perp}} u_{\parallel,s} \,+\, u_{\parallel,s} \nabla_{\parallel} u_{\parallel,s} \;=\; -\frac{q_s}{m_s} \nabla_{\parallel} \phi
\end{equation}
Assuming that electron density $n_e$ responds to electrostatic fluctuations $\phi$ according to a Boltzmann distribution, the system of equations is closed by quasi-neutrality,
\begin{equation}
	\label{eq:QN}
	\sum _{s\neq e} Z_s n_s \;=\; n_e \;=\; n_0 \, \exp{\left(\frac{e \phi}{T_e} \right)}
\end{equation}
where $T_e$ is the temperature.\\

\section{Linear analysis\label{sec:linear}}

To obtain the linear dispersion relation, densities and parallel velocities are split into equilibrium and fluctuation parts, $n_s = n_{s,0}+\tilde{n_s}$ and $u_{\parallel,s} = u_{\parallel,s0}+\tilde{u}_{\parallel,s}$. Then, Eqs.~(\ref{eq:PVG_cont_imp})-(\ref{eq:QN}) are linearized,
\begin{equation}
	\label{eq:PVG_cont_imp_lin}
	\pd{\tilde{n_s}}{t} \,+\,  \dotp{\VE}{\grad _\perp}  n_{s,0} \,+\, n_{s,0} \dotp{\grad _\perp}{\Vps} \,+\, n_{s0} \nabla_{\parallel}\tilde{u}_{\parallel,s} \;=\; 0
\end{equation}
\begin{equation}
	\label{eq:PVG_mot_imp_lin}
	\partial_t \tilde{u}_{\parallel,s} + \dotp{\VE}{\grad _\perp} u_{\parallel,s,0} \;=\; -\frac{q_s}{m_s}\nabla_{\parallel} \phi
\end{equation}
\begin{equation}
	\label{eq:QN_lin}
	\sum _{s\neq e} Z_s \tilde{n_s} \;=\; n_{e,0} \, \frac{e \phi}{T_e} 
\end{equation}
Here we used the assumption of homogeneous magnetic field to remove the term in $\dotp{\grad}{\VE}$. In addition, we neglected four terms which remain small for typical plasma parameters:
\begin{enumerate}
	\item The term $\dotp{\Vps}{\grad _\perp}  n_{s,0}$ is of order $\omega / \omega _{c,s}$ compared to $\dotp{\VE}{\grad _\perp}  n_{s,0}$. Note that, compared to $n_{s,0} \dotp{\grad _\perp}{\Vps}$, it is of order $k_x/(k_\perp^2 L_n)$, which, as verified a posteriori, remains much smaller than unity as long as $\rho_{c,i} / L_n \ll 1$.
	\item Similarly, the term $\dotp{\Vps}{\grad _\perp}  u_{\parallel,s,0}$ is of order $\omega / \omega _{c,s}$ compared to $\dotp{\VE}{\grad _\perp}  u_{\parallel,s,0}$.
	\item The term $u_{\parallel,s0} \nabla_\parallel \tilde{n_s}$, compared to $\partial \tilde{n_s} / \partial t$, is of order $k_\parallel u_{\parallel,s0}/\omega$, which must remain small to avoid strong Landau damping. Note that, compared to $\dotp{\VE}{\grad _\perp}  n_{s,0}$, it is of order $(T_i/T_e) M_s^2$, where $M_s=u_{\parallel,s0}/v_{T,i}$ is the Mach number for species $s$. To obtain the latter ordering, we substituted $k_\parallel$ by $\frac{\GradV}{2\, \overline{\omega_c}}\, k_y$ (which provides the strongest instability, cf Eq.~(\ref{eq:kpar_mostunstable}) for justification), and assumed that for each species, density and parallel gradient lengths are comparable. Extension of this theory for plasmas where the condition $M_s^2 \ll T_e/T_i$ is not satisfied, is left for future work, as accounting for the term $u_{\parallel,s0} \nabla_\parallel \tilde{n_s}$ expands a parameter space that is already very large.
	\item Similarly, the term $u_{\parallel,s0} \nabla_\parallel \tilde{u}_{\parallel,s}$, compared to $\partial \tilde{u}_{\parallel,s} / \partial t$, is of order $k_\parallel u_{\parallel,s0}/\omega$.
\end{enumerate}

A linear combination reduces the system Eqs.~(\ref{eq:PVG_cont_imp_lin})-(\ref{eq:QN_lin}) to two coupled equations, on the effective variables $\Phi = e\phi/T_e=\sum Z_s \tilde{n_s}/n_{e,0}$ and $\tilde{\bar{u}}_{\parallel} = \sum Z_s C_s \tilde{u}_{\parallel,s}$, where $C_s=n_{s,0}/n_{e,0}$ is the equilibrium concentration of species $s$. The two coupled equations write
\begin{equation}
	\partial_t \left ( 1- \bar{\rho}^2\nabla^2_{\perp} \right ) \Phi \,+\, v_{\ast,e} \, \partial_y \Phi \,+\, \nabla_{\parallel} \tilde{\bar{u}}_{\parallel}   \;=\;  0
\end{equation}
\begin{equation}
	\partial_t \tilde{\bar{u}}_{\parallel} \,-\, \GradV \, \rho_{c,i}v_{T,i}\, \partial_y \Phi 
	\;=\;  - c_s^2\nabla_{\parallel} \Phi
\end{equation}
where
\begin{equation}
	{\overline{\rho}}^2 \;=\; \sum_{s\neq e}{C_s Z_s^2 \rho_{c,s}^2}
\end{equation}
is a linear combination of ion Larmor radii,
$v_{\ast,e}=T_e /(eBL_{n,e})$ is the electron diamagnetic drift velocity,
$c_s$ is the ion-acoustic velocity including the contribution of impurities,
\begin{equation}
	c_s^2 \;=\; \sum_{s\neq e}{C_s \, Z_s^2\, \frac{T_e}{m_s}}
\end{equation}
(recalling we adopt the cold ion limit), and, most importantly,
\begin{equation}
	\GradV \;=\; \sum_{s\neq e}{C_s \, Z_s\, \nabla u_{\parallel,s,0}}
\end{equation}
is a linear combination of main ion and impurity parallel velocity gradients, which we refer to as the \emph{effective flow shear}. It is the source of free-energy of the i-PVG instability.
Note that throughout this paper, we note the radial gradient ($\partial/\partial x$) as $\nabla$ (without a subscript).
Let us define the ratio of parallel velocity gradients, $V=\nabla u_{\parallel, z, 0} / \nabla u_{\parallel, i, 0}$. Then the effective flow shear can also be written as $\GradV=(C_i + V\, C_Z\, Z_Z) \, \nabla u_{\parallel, i, 0}$.

The above model can in fact be seen as a modification of the Hasegawa-Mima model, including the effective flow shear. This analogy is however limited to the linear regime: a similar linear combination on the system Eqs.~(\ref{eq:PVG_cont_imp})-(\ref{eq:QN}) would not yield a comparable effective description of the instability at the non-linear level. However, the nonlinear evolution of a i-PVG wave packet interacting with mesoscale fluctuations (such as convective cells, zonal flows, streamers) can be investigated by modulational analysis \cite{kosuga24}. \\

Fourier analysis yields the dispersion relation,
\begin{equation}
	\omega^2 \,-\, \omega _\mathrm{iDW} \, \omega \, -\, \frac{\left(k_\parallel c_s\right)^2}{1\,+\,\left(k_\bot\overline{\rho}\right)^2}\left(1-\frac{k_y}{k_\parallel}\frac{\GradV}{\overline{\omega_c}}\right) \;=\; 0 \label{model:disprela}
\end{equation}
where
\begin{equation}
	\omega _\mathrm{iDW} \;=\; \frac{k_y \, v_{\ast,e}}{1\,+\,\left( k_\bot\overline{\rho}\right)^2}
	\label{eq:omega_iDW}
\end{equation}
is the drift-wave frequency modified by impurities, and
\begin{equation}
	\overline{\omega_c} \;=\; \sum_{s\neq e}{C_s \, Z_s \, \omega_{c,s}}
\end{equation}
is a linear combination of main ion and impurity cyclotron frequencies.\\

The quadratic Eq.~(\ref{model:disprela}) includes an unstable solution if and only if the function
\begin{equation}
	\Delta(k_{\parallel}) \;=\;  \left ( \frac{\omega_{iDW}}{2} \right )^2 + \frac{\left(k_\parallel c_s\right)^2}{1\,+\,\left(k_\bot\overline{\rho}\right)^2}\left(1-\frac{k_y}{k_\parallel}\frac{\GradV}{\overline{\omega_c}}\right)
	\label{eq:determinant}
\end{equation}
is negative.

When the parallel wave number is $k_{\parallel}=k_{\parallel,\mathrm{max}}$, with
\begin{equation}
	k_{\parallel,\mathrm{max}} \;=\; \frac{\GradV}{2\, \overline{\omega_c}}\, k_y
	\label{eq:kpar_mostunstable}
\end{equation}
the function $\Delta$ reaches its minimum value
\begin{equation}
	\Delta_{min} \;=\; \left ( \frac{\omega_{iDW}}{2} \right )^2 \, \left( 1 \,-\, L_{n,e}^2 \, (1 + k_{\perp}^2\bar{\rho}^2) \, \frac{\GradV^2}{\overline{c_s}^2} \right)
\end{equation}
Therefore, the linear instability condition can be written in terms of the effective flow shear, $\left| \GradV \right| > \GradV _{\mathrm{cr}}$ with
\begin{equation}
	\GradV _{\mathrm{cr}} ^2 \;=\; \frac{{(c_s/L_{n,e})}^2}{1+\left(k_\bot\overline{\rho}\right)^2} \label{eq:effPVGthresh}
\end{equation}
or, alternatively, in terms of the main ion flow shear, $\left| \nabla u_{\parallel,i,0}  \right| > \nabla u_{i,\mathrm{cr}}$ with
\begin{equation}
	\nabla u_{i,\mathrm{cr}} \;=\; \frac{1}{C_i + V\, C_Z\, Z_Z} \GradV _{\mathrm{cr}} \label{eq:mainionPVGthresh}
\end{equation}

When an i-PVG wave is unstable, its frequency is $\omega _\mathrm{iDW}/2$ and its growth rate is $\gamma = \sqrt{-\Delta}$ ($\Delta$ is then negative). These expressions are given in Eqs.~(\ref{eq:omega_iDW}) and (\ref{eq:determinant}). Note that the factor 2 in the frequency is mainly due to the finite value of $k_\parallel$ -- the i-PVG frequency does recover the impurity-modified drift-wave frequency in the limit of $k_\parallel = 0$ and $\GradV=0$.

For a given set of plasma parameters and a given value of $k_\perp$, the maximum growth rate is reached for $k_x=0$ and $k_\parallel=k_{\parallel,\mathrm{max}}(k_y)$, in which case
\begin{equation}
	\gamma _{max} \;=\;  \frac{k_\perp \, v_{\ast,e}/2}{1\,+\,k_\perp^2 \bar{\rho}^2}  \, \left[  L_{n,e}^2 \, (1 + k_\perp^2 \bar{\rho}^2) \, \frac{\GradV^2}{c_s^2} \,-\, 1 \right]^{1/2}  \label{eq:maximumgrowth rate}
\end{equation}
Hereafter we focus on that case $k_x=0$, which is also more consistent with our local model.

\section{Application to typical plasmas\label{sec:application}}

For concision, let us note $M=m_z/m_i$ the mass ratio, and $V=\nabla u_{\parallel, z, 0} / \nabla u_{\parallel, i, 0}$ the ratio of parallel velocity gradients. Table \ref{table:parameters} shows the parameters of a reference case, designed to be representative of a fusion plasma with highly ionized tungsten. The values $\nabla u_{\parallel, i, 0}$ and $V=2$ are consistent with measurements \cite{testa02,grierson19}.
Hereafter, we present analyses of cases which are variations of this reference case. The varying parameters are summarized in Table \ref{table:parameters2}.

\begin{table}%
	\begin{center}
		\begin{tabular}{|c|c|c|c|c|c|c|c|c|c|}
			\hline 
			& $L_n/\rho_{c,i}$ & $\nabla u_{\parallel, i, 0} / \omega_{c,i}$ & $V$ & $M$ & $Z$ & $C_z$ & $\rho_{c,i} k_x$ & $\rho_{c,i} k_y$ & $\rho_{c,i} k_\parallel$ \\ 
			\hline 
			Reference case & 10 & 0.1 & 2 & 184 & 40 & $10^{-3}$ & 0 & 0.7 & $\rho_{c,i} k_{\parallel,\mathrm{max}}$ \\ 
			\hline
		\end{tabular}
		\caption{\label{table:parameters} Input parameters for the reference case.}
	\end{center}
\end{table}

\begin{table}%
	\begin{center}
		\begin{tabular}{|c|c|c|c|c|c|c|c|}
			\hline 
			&  $\nabla u_{\parallel, i, 0} / \omega_{c,i}$ & $V$ & $M$ & $Z$ & $C_z$  & $\rho_{c,i} k_y$ & $\rho_{c,i} k_\parallel$ \\ 
			\hline 
			Fig.~\ref{fig:disprealkz} & 0.1 & 2 & 12 and 184 & 6 and 40 & 0 and $10^{-3}$ &  0.7 & -0.02 -- 0.08\\ 
			\hline 
			Fig.~\ref{fig:disprealkzky} &  0.1 & 2 & 184 & 40 & $10^{-3}$ &  -1 -- 1 & -0.1 -- 0.1\\ 
			\hline 
			Fig.~\ref{fig:disprealky} &  0.1 & 2 & 12 and 184 & 6 and 40 & 0 and $10^{-3}$ &  0 -- 2 & $\rho_{c,i} k_{\parallel,\mathrm{max}}$\\ 
			\hline 
			Fig.~\ref{fig:typedep_dvi} &  - & 2 & 0 -- 200 & 0 -- 200 & $10^{-3}$ &  0.7 & $\rho_{c,i} k_{\parallel,\mathrm{max}}$\\ 
			\hline 
			Fig.~\ref{fig:M_dvi} &  0.1 & -1 -- 2 & 5 -- 200 & 5 & $10^{-4}$ -- $10^{-2}$ &  0.7 & $\rho_{c,i} k_{\parallel,\mathrm{max}}$\\ 
			\hline 
			Fig.~\ref{fig:Z_dvi} &  0.1 & -1 -- 2 & 184 & 0 -- 50 & $10^{-4}$ -- $10^{-2}$ &  0.7 & $\rho_{c,i} k_{\parallel,\mathrm{max}}$\\ 
			\hline 
			Fig.~\ref{fig:Cz_dvi} &  0.1 & -1 -- 2 & 12 -- 184 & 3 -- 40 & $10^{-5}$ -- $10^{-2}$ &  0.7 & $\rho_{c,i} k_{\parallel,\mathrm{max}}$\\ 
			\hline 
			Fig.~\ref{fig:disprealky_threshold} &  0.08 & 2 &  12 and 184 & 6 and 40 & 0 and $10^{-3}$ &  0 -- 2 & $\rho_{c,i} k_{\parallel,\mathrm{max}}$\\ 
			\hline 
			Fig.~\ref{fig:typedep_gamma} &  0.08 & 2 &  0 -- 200 & 0 -- 200 & $10^{-3}$ &  0.7 & $\rho_{c,i} k_{\parallel,\mathrm{max}}$\\ 
			\hline 
			Fig.~\ref{fig:M_gamma} &  0.08 & -1 -- 2 & 5 -- 200 & 5 & $10^{-4}$ -- $10^{-2}$ &  0.7 & $\rho_{c,i} k_{\parallel,\mathrm{max}}$\\ 
			\hline 
			Fig.~\ref{fig:Z_gamma} &  0.08 & -1 -- 2 & 184 & 0 -- 50 & $10^{-4}$ -- $10^{-2}$ &  0.7 & $\rho_{c,i} k_{\parallel,\mathrm{max}}$\\ 
			\hline 
			Fig.~\ref{fig:Cz_gamma} &  0.08 & -1 -- 2 & 12 -- 184 & 3 -- 40  & $10^{-5}$ -- $10^{-2}$ &  0.7 & $\rho_{c,i} k_{\parallel,\mathrm{max}}$\\ 
			\hline
		\end{tabular}
		\caption{\label{table:parameters2} Input parameters for each figure. This table includes only parameters which vary compared to the reference case. In other words, for all figures, $L_n/\rho_{c,i}=10$ and $k_x = 0$.}
	\end{center}
\end{table}

\newpage
\subsection{Dispersion relation}

\begin{figure}[!htb]
	\begin{center}
		\includegraphics[width=0.6\textwidth]{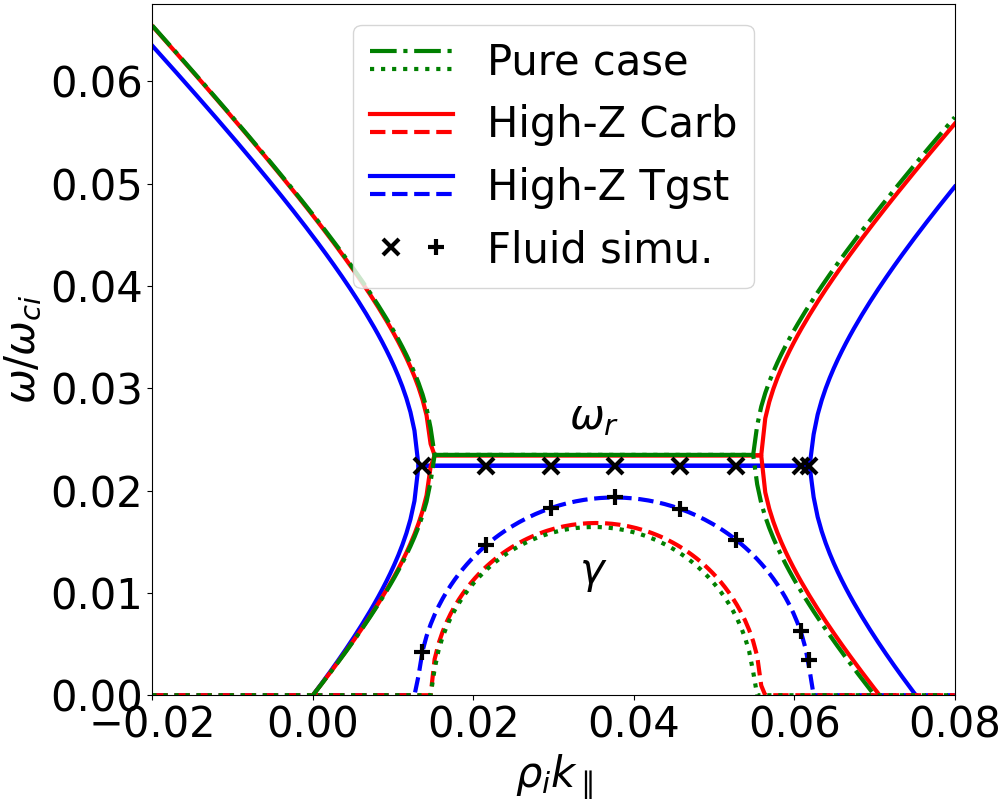}
		\caption{\label{fig:disprealkz} Dispersion relation of the i-PVG: frequency and growth rate against parallel wavenumber, for the reference case ("High-Z Tgst"), as well as a case with C$^{6+}$ impurity ("High-Z Carb"). The dispersion relation of the pure PVG is included for comparison. The curves for the pure PVG are almost hidden behind the curves for the C$^{6+}$ impurity. The cross and plus symbols correspond to linear frequency and growth rate measured in an initial value simulation of the fluid model Eqs.~(\ref{eq:PVG_cont_imp}-\ref{eq:QN}).}
\end{center} \end{figure}

Figure \ref{fig:disprealkz} shows the dispersion relation for the reference case, as well as for C$^{6+}$ ($M=12$ and $Z=6$) impurity. The dispersion relation of the pure PVG is included for comparison. We observe that the dispersion relation is qualitatively similar for the pure and impure cases, but quantitatively different.
As demonstrated hereafter, differences are more important closer to the PVG instability threshold, or for higher impurity concentrations.

Let us now verify both the accuracy of the assumptions (i)--(iv) described in Sec.~\ref{sec:linear}, and the correctness of our linear analysis. We developed an initial value numerical simulation code to solve the fluid model Eqs.~(\ref{eq:PVG_cont_imp}-\ref{eq:QN}), which does not rely on the assumptions (i)--(iv). Time integration is performed by an explicit RK4 scheme with time-step width $\Delta t = 10^{-2} \omega _{c,i}^{-1}$. All quantities ($\phi$, $n_s$, $u_{\parallel, s}$) are described in Fourier space ($k_x$, $k_y$) on a 256x256 grid for a fixed $k_\parallel$. Nonlinear terms such as $n_s \, u_{\parallel, s}$ are computed in real space before being transformed back to Fourier space.
Fig.~\ref{fig:disprealkz} includes the frequency and growth rate extracted from time-traces of complex Fourier component of $\phi$, during the linear phase. There is good quantitative agreement for the values of $k_\parallel$ such that the growth rate is close to its maximum -- the relative error for the growth rate remains below $0.2\%$ for the range $\gamma / \omega _{c,i}> 0.014 $. The relative error can be as high as a few percent in the range $0.004<\gamma / \omega _{c,i}<0.007$ or a few tens of percent in the range $\gamma / \omega _{c,i}<0.004$, but these weak modes are negligible in the linear phase of the instability. The relative error for the frequency remains below $0.5\%$ for the whole unstable range.

\begin{figure}[!htb]
	\begin{center}
		\includegraphics[width=0.6\textwidth]{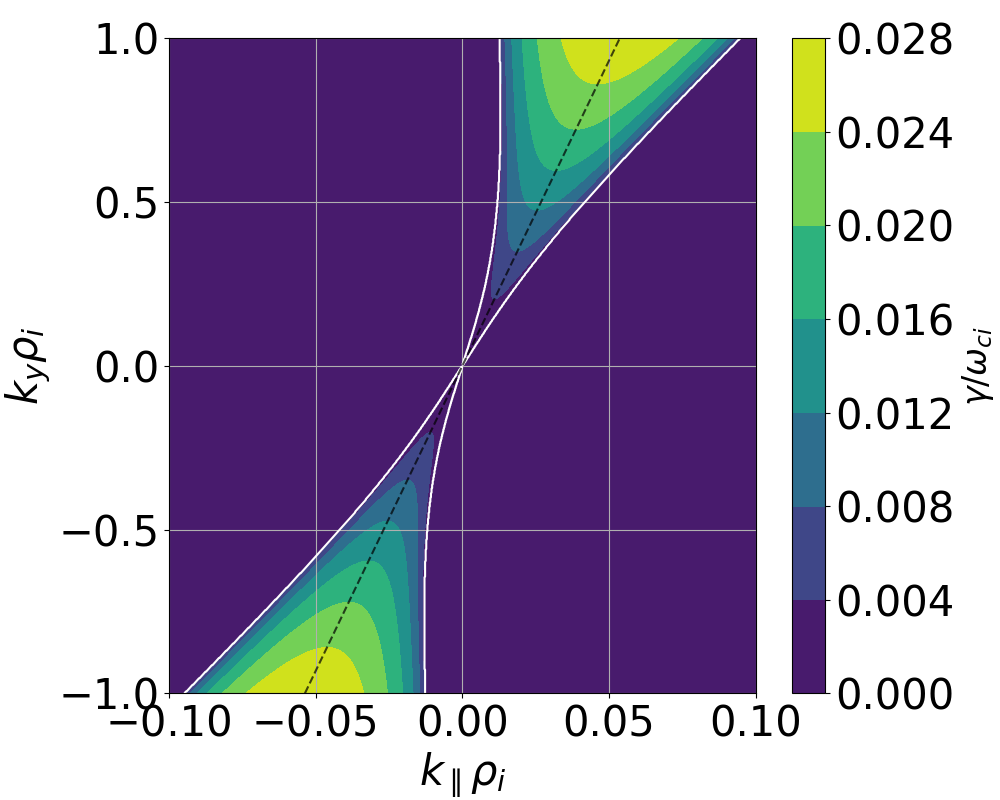}
		\caption{\label{fig:disprealkzky} Growth rate against parallel and azimuthal wavenumbers, for the reference case. The dashed black line represents the condition $k_\parallel = k_{\parallel,\mathrm{max}}$. The white curve is the instability threshold.}
\end{center} \end{figure}

Figure \ref{fig:disprealkzky} shows the linear growth rate in the space of wavevector components ($k_\parallel$, $k_y$).
The asymmetry in this space, instability occurring only for $k_y \, k_\parallel > 0$ (for $\GradV>0$), is an important property of the i-PVG. If the sign of $\GradV$ was flipped, the unstable region would be in the region $k_y \, k_\parallel < 0$.
As expected, growth rate is maximum for $k_\parallel = k_{\parallel,\mathrm{max}}$. Hereafter, we always set $k_{\parallel} = k_{\parallel,\mathrm{max}}$.

\begin{figure}[!htb]
	\begin{center}
		\includegraphics[width=0.6\textwidth]{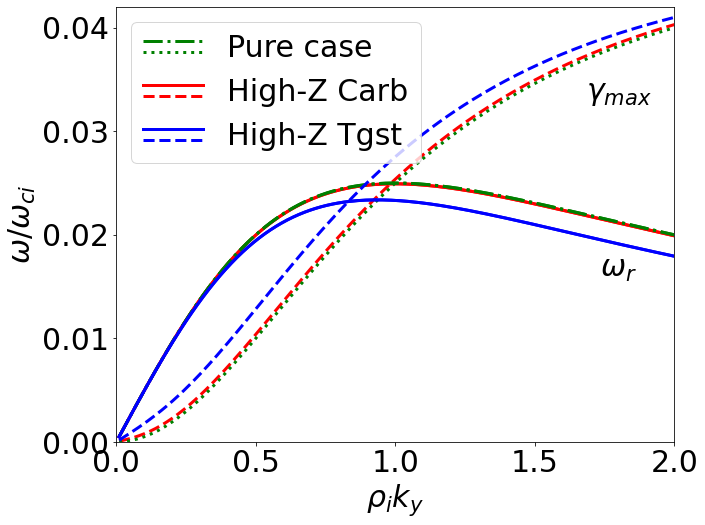}
		\caption{\label{fig:disprealky} Dispersion relation of the i-PVG: frequency and growth rate against the azimuthal wavenumber $k_y$, for the reference case ("High-Z Tgst"), as well as a case with C$^{6+}$ impurity ("High-Z Carb"). The dispersion relation of the pure PVG is included for comparison. The curves for the pure PVG are almost hidden behind the curves for the C$^{6+}$ impurity.}
\end{center} \end{figure}

\newpage 
Figure \ref{fig:disprealky} displays the maximum growth rate (obtained for $k_\parallel=k_{\parallel,\mathrm{max}}$) against the azimuthal $k_y$ wavenumber, as well as the corresponding frequency. Our model does not include any mechanism for small scale dissipation. As a consequence, $\gamma_\mathrm{max}$ spuriously keeps on increasing for increasing $k_y$. However, our model looses its validity as $\rho _{c,i} k$ approaches unity.
With the knowledge of the typical shape of $\gamma_\mathrm{max}(k)$ obtained from more complete models for many instabilities, let us approximate the physical maximum growth rate by choosing a cut-off at $\rho _{c,i} k = 0.7$ as a first, qualitative approach. Then, for our reference case, the maximum growth rate is $\gamma_\mathrm{max} \sim 0.02\,  \omega_{ci}$.

In summary, the results in this section confirm that the linear properties of the i-PVG instability are qualitatively the same as that of the pure PVG instability. For the chosen set of parameters, even the quantitative differences are small. However, that is for a case where the pure PVG instability is already quite strong, since $\gamma \approx \omega$. In this sense, the above results concern a range far from instability threshold. By contrast, in the next section we focus on the instability threshold itself.

\subsection{Critical ionic PVG}

As explained in Sec.~\ref{sec:linear}, for a given set of parameters ($L_n$, $\nabla u_{\parallel, i, 0}$, $V$, $M$, $Z$, $C_z$), and bounded value of $k_y$, the i-PVG instability threshold can be expressed either in terms of the effective flow shear, which is a combination of main ion and impurity flow shears, or, alternatively, in terms of the main ion flow shear alone. The corresponding thresholds are
given by Eqs.~(\ref{eq:effPVGthresh}) and (\ref{eq:mainionPVGthresh}), respectively. The preferred point-of-view may depend on the experimental conditions which determine preferred control parameters. Here we focus on the second point-of-view, that of critical main ion flow shear for a given ratio $V$.

\begin{figure}%
	\begin{center}
		\includegraphics[width=0.6\textwidth]{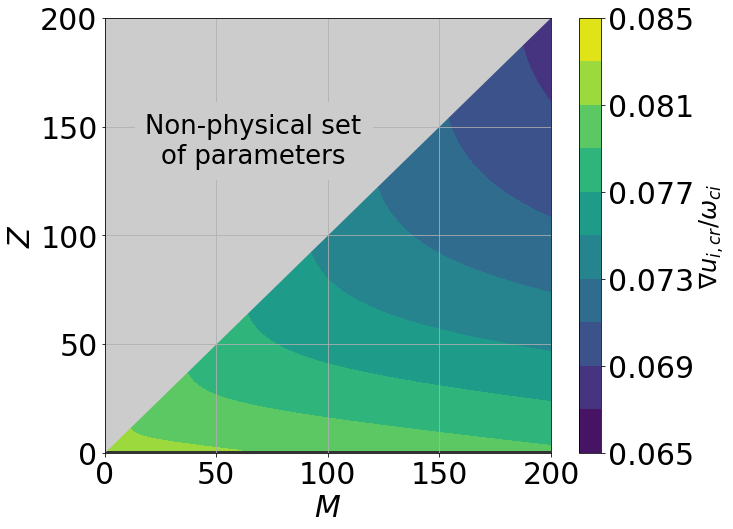}
		\caption{\label{fig:typedep_dvi} Critical main ion flow shear against mass and charge ratios.}
\end{center} \end{figure}

Figure \ref{fig:typedep_dvi} displays the critical main ion flow shear against mass and charge ratios, for $V=2$ and $C_z=10^{-3}$. It is clear that i-PVG-unstable conditions are more readily reached when the impurity is massive and highly ionized. In realistic machines however, radiation from high-mass and high-Z impurities is a severe issue, so that the applicability of such ranges of parameters should be carefully studied. On the contrary, at this low concentration, the threshold for low-mass, low-charge impurities is not so different from the threshold for high-mass, low-charge impurities, and is actually close to the threshold of the pure PVG.

\begin{figure}[!htb]
	\begin{center}
		\includegraphics[width=0.45\textwidth]{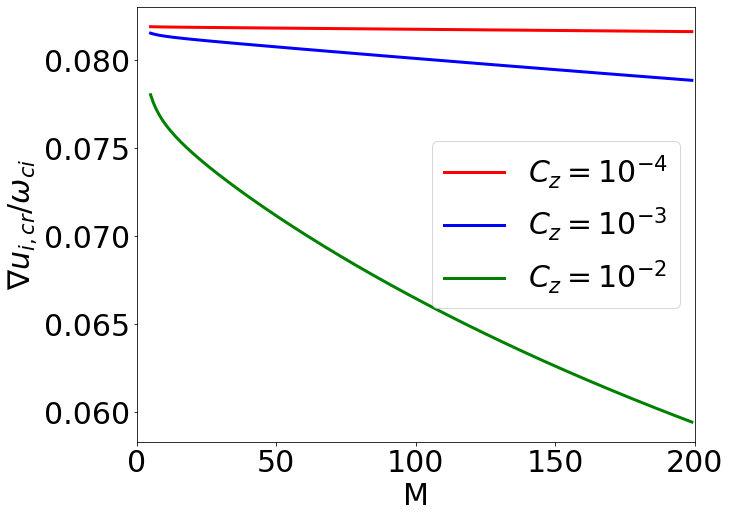}
		\includegraphics[width=0.45\textwidth]{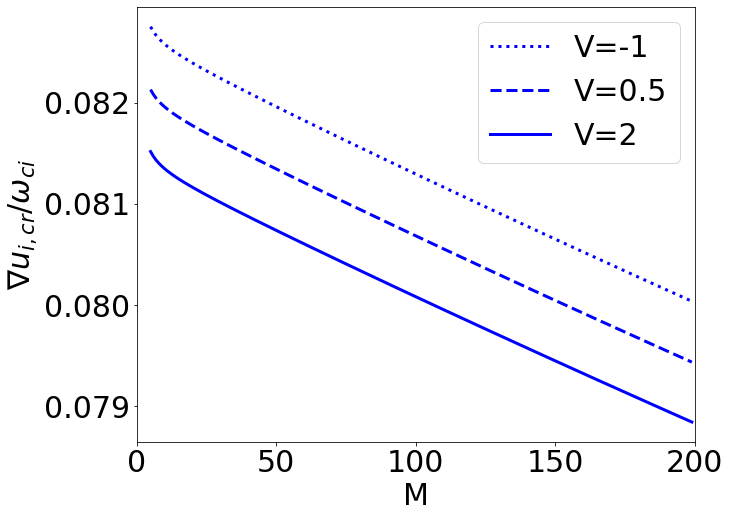}
		\caption{\label{fig:M_dvi} Critical main ion flow shear against mass ratio, for $Z=5$. Left: fixed $V=2$ and various concentrations. Right: fixed $C_z=10^{-3}$ and various $V$.}
\end{center} \end{figure}

\newpage
Figure \ref{fig:M_dvi} illustrates the role of the mass ratio, at fixed $Z = 5$, for various impurity concentrations $C_z$ and various flow shear ratios $V$. The critical main ion flow shear is always decreasing with increasing mass ratio, but the relationship is nonlinear. Here the impact of $V$ is small, but that is because $C_z$ and $Z$ are both low, such that the i-PVG is actually very close to the pure PVG.\\

\begin{figure}[!htb]
	\begin{center}
		\includegraphics[width=0.45\textwidth]{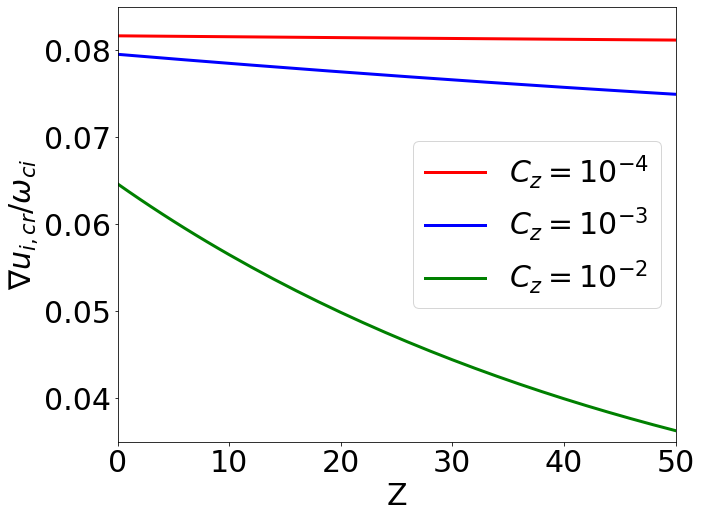}
		\includegraphics[width=0.45\textwidth]{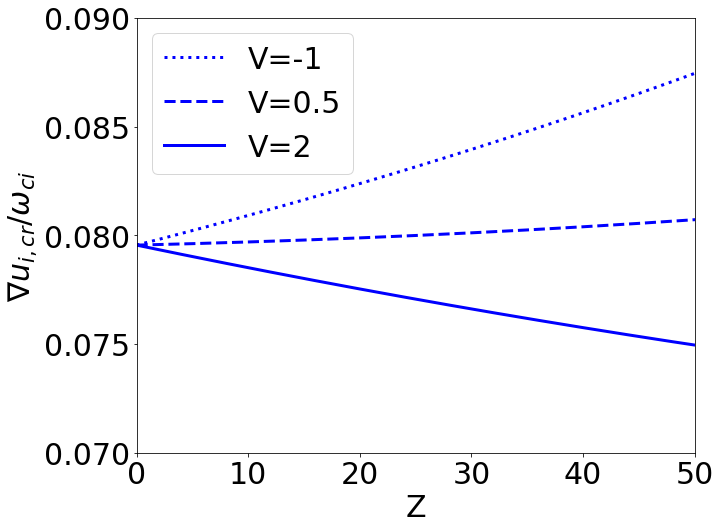}
		\caption{\label{fig:Z_dvi} Critical main ion flow shear against charge number $Z$, for $M=184$. Left: fixed $V=2$ and various concentrations. Right: fixed $C_z=10^{-3}$ and various $V$.}
\end{center} \end{figure}

Figure \ref{fig:Z_dvi} illustrates the role of the charge number $Z$ of the impurity, at fixed mass ratio $M=184$ (tungsten), for various impurity concentrations $C_z$ and various flow shear ratios $V$. Higher $Z$ can be either stabilising or destabilising depending on the parameters, such as $V$. Here, at concentration $C_z=10^{-3}$, it is stabilising for $V=2$, but destabilising for $V=0.5$ and $V=-1$.
This stabilising role of impurities is consistent with simple intuition in the case $V=-1$, since impurity flow counteracts main ion flow in this case. In the case $V=0.5$, the stabilising effect is more subtle: in this case the variation in Eq.~(\ref{eq:mainionPVGthresh}) is dominated by the denominator $(C_i+V C_Z Z)$, which can be re-written as $[1+Z C_Z (V-1)]$. However, which term dominates depends on the value of $V$. For example, for $V=1$, the variation as $Z$ increases is dominated by the decrease of $c_s$, so increasing $Z$ is stabilising. For an intermediate case, $V=0.8$, the dependency is non-monotonous: increasing $Z$ is stabilising for $Z<28$ and destabilising for $Z\geq 28$.\\

\begin{figure}[!htb]
	\begin{center}
		\includegraphics[width=0.45\textwidth]{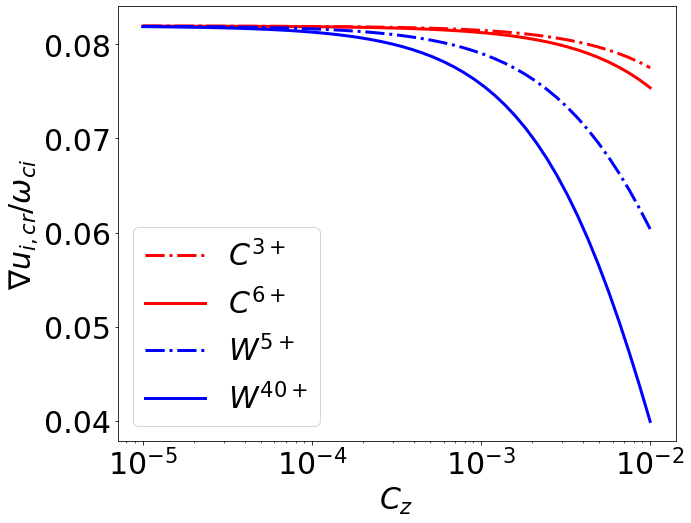}
		\includegraphics[width=0.45\textwidth]{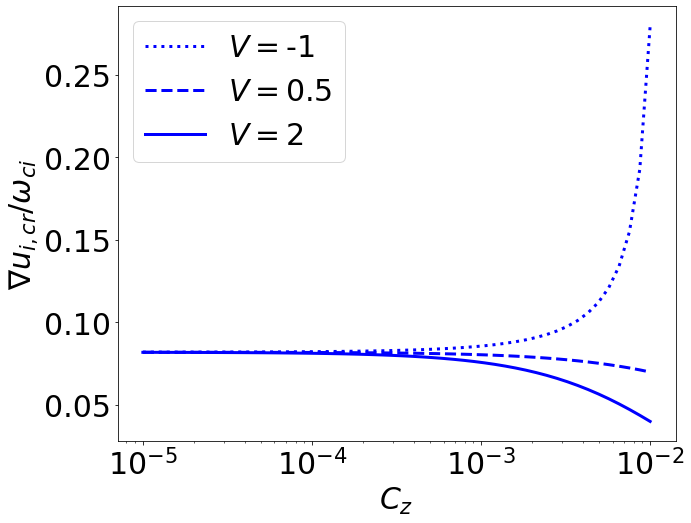}
		\caption{\label{fig:Cz_dvi} Critical main ion flow shear against impurity concentration. Left: fixed $V=2$ and various species. Right: various $V$ and the impurity is W$^{40+}$.}
\end{center} \end{figure}

A more intuitive picture can be obtained by varying impurity concentration while the impurity species and $Z$ are fixed.
Figure \ref{fig:Cz_dvi} illustrates the role of impurity concentration, for various impurity species (C$^{3+}$, C$^{6+}$, W$^{5+}$, W$^{40+}$) and various values of $V$. From this point of view, the role of impurity flow is consistent with simple intuition: if $V>0$, impurity flow adds up to main ion flow, which is destabilising; conversely, if $V<0$, impurity flow counters main ion flow, which is stabilising. Increasing concentration accentuates the impact -- whether stabilising or destabilising -- of the impurity on the instability threshold.
This effect may offer new means of control of the PVG instability by impurity injection.\\

In summary, the critical main ion flow shear decreases non-linearly with increasing mass ratio, and may increase or decrease in a non-trivial manner depending on all the other parameters.

For our reference case, the instability threshold is $\nabla u_{i,\mathrm{cr}} = 0.075925 \, \omega_{c,i}$.
In the next subsection, in order to study the growth rate near the instability threshold, we arbitrarily fix $\nabla u_{\parallel,i,0} = 0.08 \, \omega_{c,i}$.

\subsection{Growth rate}

\begin{figure}[!htb]%
	\begin{center}
		\includegraphics[width=0.6\textwidth]{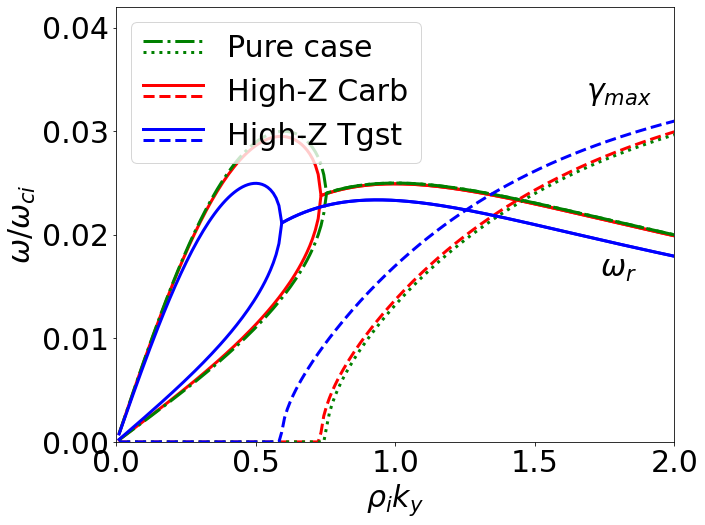}
		\caption{\label{fig:disprealky_threshold} Dispersion relation of the i-PVG: frequency and growth rate against azimuthal wavenumber, for the reference case, as well as a case with C$^{6+}$ impurity ("High-Z Carb") -- same as Fig.~\ref{fig:disprealkz} but for lower $\nabla u_{\parallel,i,0} = 0.08 \, \omega_{c,i}$. The dispersion relation of the pure PVG is included for comparison. The curves for the pure PVG are almost hidden behind the curves for the C$^{6+}$ impurity.}
\end{center} \end{figure}

Figure \ref{fig:disprealky_threshold} displays the maximum growth rate (obtained for $k_\parallel=k_{\parallel,\mathrm{max}}$) against the azimuthal $k_y$ wavenumber, as well as the corresponding frequency, for $\nabla u_{\parallel,i,0} = 0.08 \, \omega_{c,i}$. Here, for $k_y<0.75 \, \rho_{c,i}^{-1}$, the main ion parallel velocity shear is below the threshold for pure PVG.
As a consequence, the behavior is qualitatively different compared to Fig.~\ref{fig:disprealkz} (which is for $\nabla u_{\parallel,i,0} = 0.1 \, \omega_{c,i}$, above the pure PVG threshold even for $k_y \rightarrow 0$). There is now an additional bifurcation.\\

\begin{figure}[!htb]
	\begin{center}
		\includegraphics[width=0.45\textwidth]{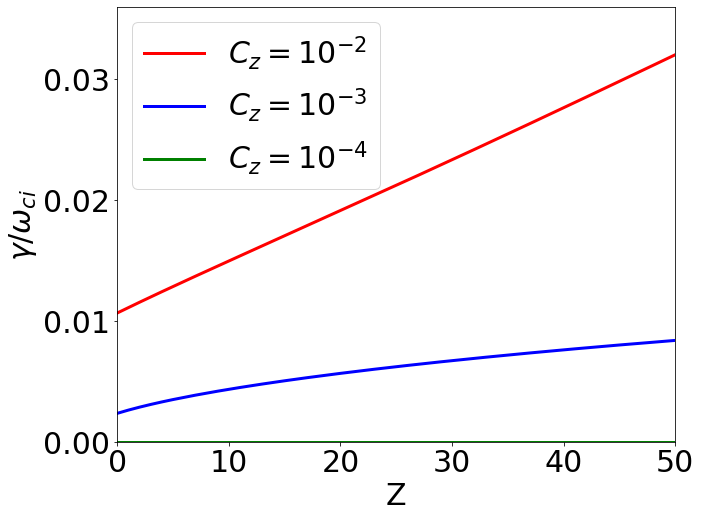}
		\includegraphics[width=0.45\textwidth]{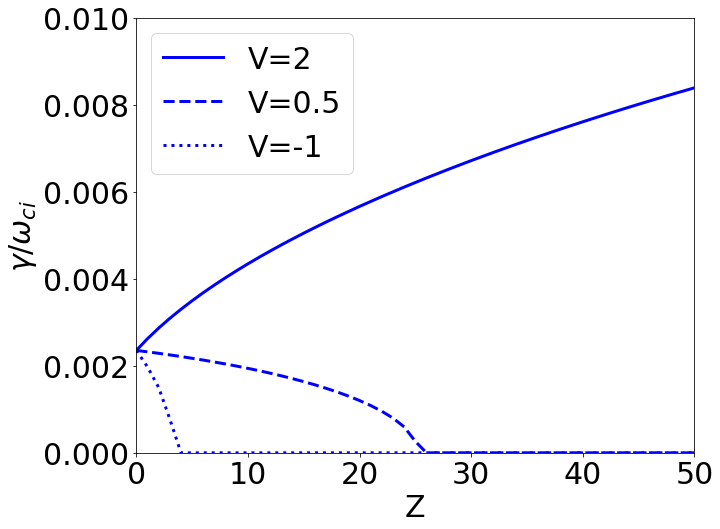}
		\caption{\label{fig:Z_gamma} Maximum growth rate against impurity charge, for $\nabla u_{\parallel,i,0} = 0.08 \, \omega_{c,i}$ and $M=184$. Left: fixed $V=2$ and various concentrations. Right: fixed $C_z=10^{-3}$ and various $V$.}
\end{center} \end{figure}

\begin{figure}[!htb]
	\begin{center}
		\includegraphics[width=0.45\textwidth]{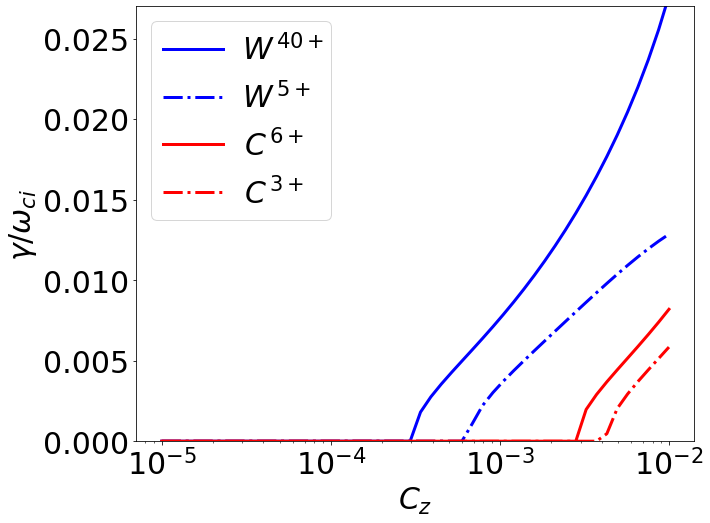}
		\includegraphics[width=0.45\textwidth]{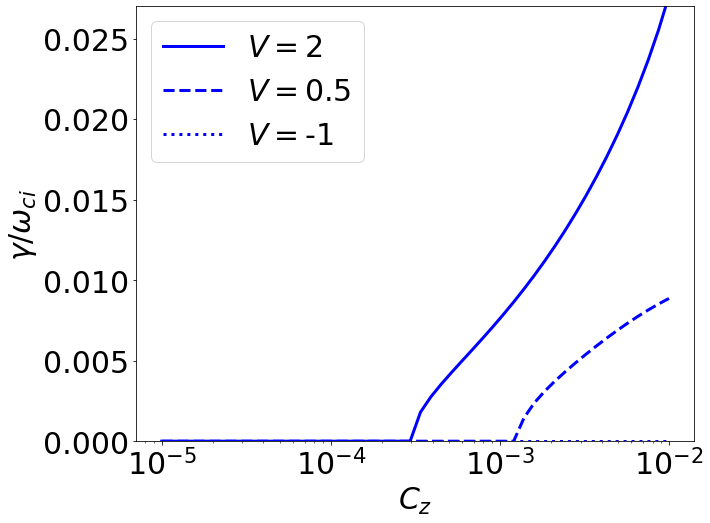}
		\caption{\label{fig:Cz_gamma} Maximum growth rate against impurity concentration. Left: fixed $V=2$ and various species. Right: various $V$ and the impurity is W$^{40+}$.}
\end{center} \end{figure}

Again, let us approximate the physical maximum growth rate by fixing $\rho _{c,i} k = 0.7$. Then, for our plasma with $W^{40+}$, the maximum growth rate is $\gamma_\mathrm{max} \sim 0.008\,  \omega_{c,i}$.
Consistently, the maximum growth rate in the following figures (\ref{fig:typedep_gamma}, \ref{fig:M_gamma}, \ref{fig:Z_gamma} and \ref{fig:Cz_gamma}), where impurity parameters are varied, is obtained from our model for the wave vector $\vec{k}_\mathrm{max}$ with components $k_x = 0$, $k_y = 0.7 / \rho _{c,i}$, $k_{\parallel}=k_{\parallel,\mathrm{max}}$.

Figure \ref{fig:typedep_gamma} shows our estimated maximum growth rate against mass and charge ratios, for $V=2$ and $C_z=10^{-3}$. Recalling that the real frequency of the mode is typically $\omega \approx 0.02 \, \omega_{c,i}$, we confirm that our choice of cut-off ($\rho _{c,i} k = 0.7$) ensures that $\gamma _\mathrm{max}/\omega$ remains significantly below unity, except for unrealistically highly ionized heavy impurities.
Note that here, the main ion parallel velocity shear is below the threshold for pure PVG (which is $0.08192 \, \omega_{c,i}$ for $k_y=0.7 \, \rho_{c,i}^{-1}$).

\begin{figure}[!htb]
	\begin{center}
		\includegraphics[width=0.6\textwidth]{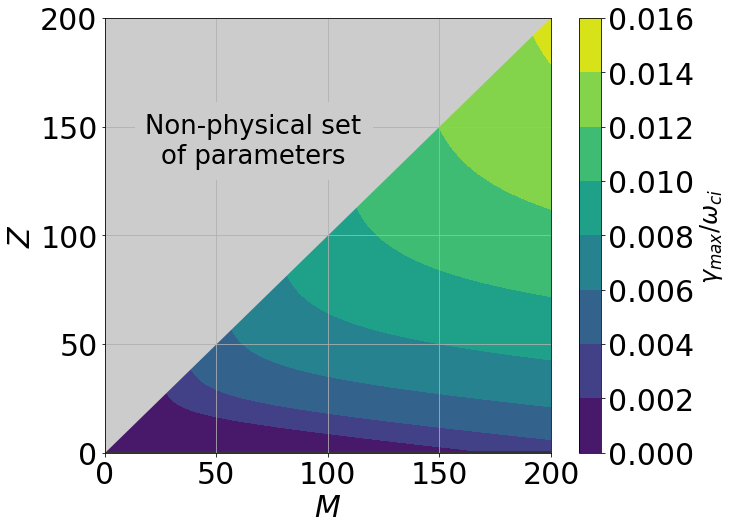}
		\caption{\label{fig:typedep_gamma} Maximum growth rate against mass and charge ratios, for $\nabla u_{\parallel,i,0} = 0.08 \, \omega_{c,i}$.}
\end{center} \end{figure}

\begin{figure}[!htb]
	\begin{center}
		\includegraphics[width=0.45\textwidth]{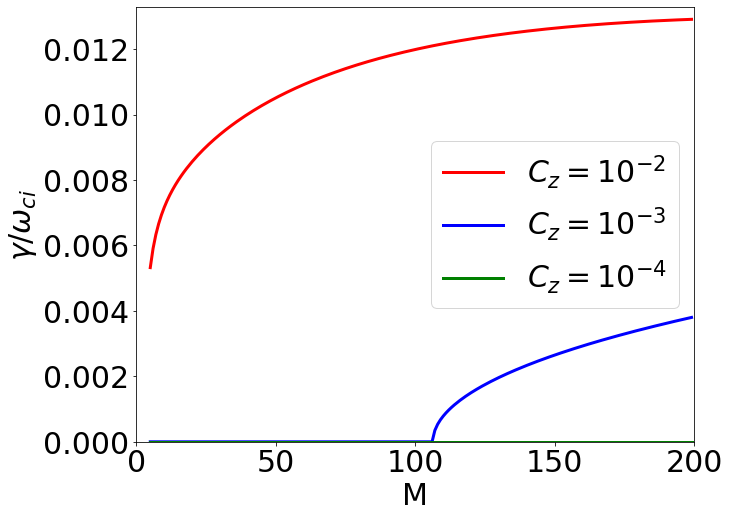}
		\includegraphics[width=0.45\textwidth]{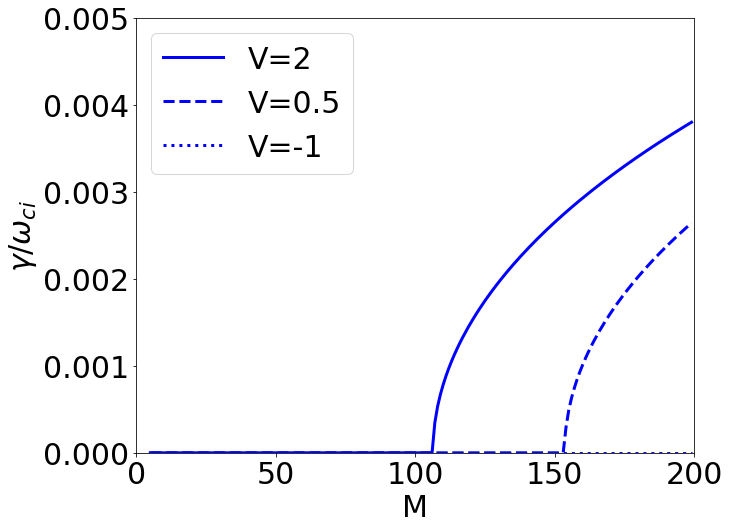}
		\caption{\label{fig:M_gamma} Maximum growth rate against mass ratio, for $\nabla u_{\parallel,i,0} = 0.08 \, \omega_{c,i}$ and $Z=5$. Left: fixed $V=2$ and various $C_z$. Right: fixed $C_z=10^{-3}$ and various $V$.}
\end{center} \end{figure}

Figure \ref{fig:M_gamma} illustrates the role of the mass ratio on the maximum growth rate, for fixed $Z = 5$ and various $C_z$ and $V$. From this perspective, the instability threshold can be viewed as a threshold in impurity mass, every other parameters being fixed. Similarly to the effect of mass on the instability threshold, increasing mass always increases the maximum growth rate.

\newpage
Figure \ref{fig:Z_gamma} illustrates the role of the impurity charge on the maximum growth rate, for fixed $M=184$ and various $C_z$ and $V$. As a rule of thumb, the growth rate is less sensitive to $Z$ than to $M$. Increasing $Z$ may be destabilising or stabilising depending on the parameters, and in particular depending on the flow shear ratio $V$. For $V=0.5$ or $V=-1$, the i-PVG instability threshold can be viewed as a threshold in impurity charge, instability occurring only \emph{below} a critical number.

Figure \ref{fig:Cz_gamma} illustrates the role of impurity concentration on the maximum growth rate, for various species and $V$. In this point-of-view, the instability threshold is a threshold in concentration. With our arbitrary choice of $\nabla u_{\parallel,i,0} = 0.08 \, \omega_{c,i} = 1.05 \nabla u_{i,\mathrm{cr}}$, the concentration threshold is of the order of $10^{-4}$ -- $10^{-3}$ for tungsten, and of the order of $10^{-3}$ -- $10^{-2}$ for carbon. However, with $\nabla u_{\parallel,i,0}$ closer to $\nabla u_{i,\mathrm{cr}}$, the concentration threshold can even be much lower.
Here, the growth rate increases with increasing concentration. However, there are also cases where concentration has a non-monotonous impact \cite{lesur23iaea}.

\newpage
Let us emphasize that figures \ref{fig:Z_gamma} and \ref{fig:Cz_gamma} are useful to guide experiments which could test this theory. For example, impurities may be purposely injected in a linear device, to either excite the i-PVG, or to suppress a pure PVG (depending on the direction of flow shears). However, the analysis should be reproduced after measuring the values of parallel flows, since the thresholds in $Z$ and $C_Z$ are sensitive to the value of the main ion flow shear -- and more specifically its closeness to marginal stability.

\section{Conclusions and discussions}

We investigated the role of a radial gradient of the parallel flow of impurities in a magnetized plasma. The linear dispersion relation indicates that the impurity-PVG is destabilised when $ L_{n,e} ^2 \GradV ^2 > c_s^2/[1+\left(k_\bot\overline{\rho}\right)^2]$.
Similarly to the pure-PVG, the i-PVG is asymmetric in the space of wavevector components ($k_\parallel$, $k_y$): it is unstable only for $k_y \, k_\parallel \GradV > 0$.

If the plasma is unstable to the pure-PVG, then impurities can be stabilising if their flow shear is opposite to the flow shear of main ions. Note that this does not require flows themselves to be in opposite directions.
The role of impurities is more prominent for higher mass and/or higher charge number. In general, the instability threshold and the growth rate are more sensitive to charge number than to mass.

The analytic expression of the maximum growth rate is given in Eq.~(\ref{eq:maximumgrowth rate}).
To investigate how the maximum growth rate depends on plasma and impurity parameters (Figs.~\ref{fig:M_gamma}-\ref{fig:Cz_gamma}), we assumed a perpendicular wavenumber comparable to the inverse Larmor radius (of main ions).

In the typical case of a hydrogen plasma with a $10^{-3}$ concentration of $W^{40+}$, tungsten flow shear being twice the hydrogen flow shear, and assuming that the linear growth rate is maximum for $k_\perp \approx 0.7 \, \rho_{c,i}^{-1}$, the i-PVG instability is unstable for hydrogen flow shear above a threshold $\nabla u_{i,\mathrm{cr}} \approx 0.076 \, \omega_{c,i}$, and most unstable for parallel wavenumbers $k_\parallel \approx 0.04 \, \rho_{c,i}^{-1}$.

Although the present work should in principle capture the main linear properties of the i-PVG, the reader should keep in mind that it is subject to several caveats: it is a local analysis, assuming cold ions and Boltzmann electrons. Our model does not include collisional effects, viscosity, resistivity. We limited our analysis to two species, but it can be straightforwardly generalised to more species.

The present work focused on the linear properties of the i-PVG. Although, the (quasi-)steady-state spectrum of fluctuations may roughly reflect the location in $\vec{k}$-space of most unstable modes, nonlinear simulations and analysis are necessary to improve predictions and understanding of steady-state fluctuations. In particular, the present work cannot address the role of zonal flows, which can be generated by PVGs \cite{kosuga17pop}. Impurities may play an interesting role in both the generation and the damping of zonal flows.

Impurities are transported in the radial direction, in general by both neoclassical and turbulent processes, although one process may dominate depending on the species and plasma parameters. The efficiency of fusion depends on the radial distribution of impurities, which results from transport. For example, slight tungsten contamination of the core yields prohibitive energy losses. Therefore, it is essential to improve our understanding of impurity transport.
However, the model adopted here assumes a Boltzmann response of electrons to electric fluctuations, which precludes any radial transport of electrons. Quasi-neutrality then forces impurity particle transport to exactly balance main ion particle transport. This assumption is reasonable for the present linear analysis, but is too strong for providing any conclusion about impurity particle transport. The relaxation of this assumption is left for future work. In addition, we recall that our model assumes cold ions, which precludes us from making any similar statement about heat transport.


\section*{Acknowledgments}
\addcontentsline{toc}{section}{Acknowledgments}

This work was supported by a scholarship by Kyushu University, the joint research project in RIAM, the Grants-in-Aide for Scientific Research of JSPS of Japan (JP21H01066, JP23K20838), a grant from the Chaire Energies Durables of Ecole Polytechnique, and funding from the Agence Nationale de la Recherche for the project GRANUL (ANR-19-CE30-0005).\\


\bibliographystyle{aomalpha}
\bibliography{tout}

\providecommand{\etalchar}[1]{$^{#1}$}
\providecommand{\bysame}{\leavevmode\hbox to3em{\hrulefill}\thinspace}
\providecommand{\noopsort}[1]{}
\providecommand{\mr}[1]{\href{http://www.ams.org/mathscinet-getitem?mr=#1}{MR~#1}}
\providecommand{\zbl}[1]{\href{http://www.zentralblatt-math.org/zmath/en/search/?q=an:#1}{Zbl~#1}}
\providecommand{\jfm}[1]{\href{http://www.emis.de/cgi-bin/JFM-item?#1}{JFM~#1}}
\providecommand{\arxiv}[1]{\href{http://www.arxiv.org/abs/#1}{arXiv~#1}}
\providecommand{\doi}[1]{\url{https://doi.org/#1}}
\providecommand{\MR}{\relax\ifhmode\unskip\space\fi MR }
\providecommand{\MRhref}[2]{%
  \href{http://www.ams.org/mathscinet-getitem?mr=#1}{#2}
}
\providecommand{\href}[2]{#2}
\begin{thebibliography}{WWY{\etalchar{+}}98}

\bibitem[BPH{\etalchar{+}}11]{barnes11}
\bgroup\scshape{}M.~Barnes\egroup{}, \bgroup\scshape{}F.~Parra\egroup{},
  \bgroup\scshape{}E.~Highcock\egroup{},
  \bgroup\scshape{}A.~Schekochihin\egroup{},
  \bgroup\scshape{}S.~Cowley\egroup{}, and \bgroup\scshape{}C.~Roach\egroup{},
  Turbulent transport in tokamak plasmas with rotational shear,  \emph{Phys.
  Rev. Lett.} \textbf{106} no.~17 (2011), 175004.
  \doi{10.1103/PhysRevLett.106.175004}.

\bibitem[BGH{\etalchar{+}}23]{brochard2023spektre}
\bgroup\scshape{}F.~Brochard\egroup{}, \bgroup\scshape{}D.~Geneve\egroup{},
  \bgroup\scshape{}S.~Heuraux\egroup{}, \bgroup\scshape{}V.~Bobkov\egroup{},
  \bgroup\scshape{}D.~Del~Sarto\egroup{}, \bgroup\scshape{}E.~Faudot\egroup{},
  \bgroup\scshape{}A.~Ghizzo\egroup{}, \bgroup\scshape{}E.~Gravier\egroup{},
  \bgroup\scshape{}N.~Lemoine\egroup{}, \bgroup\scshape{}M.~Lesur\egroup{},
  \bgroup\scshape{}N.~Louis\egroup{}, \bgroup\scshape{}J.~Moritz\egroup{},
  \bgroup\scshape{}T.~Reveillé\egroup{}, \bgroup\scshape{}V.~Rohde\egroup{},
  \bgroup\scshape{}U.~Stroth\egroup{}, \bgroup\scshape{}G.~Urbanczyk\egroup{},
  \bgroup\scshape{}F.~Volpe\egroup{}, and \bgroup\scshape{}H.~Zohm\egroup{},
  Spektre, a linear radiofrequency device for investigating edge plasma
  physics,  in \emph{49th EPS Conference on Plasma Physics}, IAEA Publications,
  Vienna, 2023, p.~THS/P5–03.

\bibitem[CRL73]{catto73}
\bgroup\scshape{}P.~J. Catto\egroup{}, \bgroup\scshape{}M.~N.
  Rosenbluth\egroup{}, and \bgroup\scshape{}C.~Liu\egroup{}, Parallel velocity
  shear instabilities in an inhomogeneous plasma with a sheared magnetic field,
   \emph{Phys. Fluids} \textbf{16} no.~10 (1973), 1719--1729.
  \doi{10.1063/1.1694200}.

\bibitem[CBKW12]{chapman12}
\bgroup\scshape{}I.~Chapman\egroup{}, \bgroup\scshape{}S.~Brown\egroup{},
  \bgroup\scshape{}R.~Kemp\egroup{}, and \bgroup\scshape{}N.~Walkden\egroup{},
  Toroidal velocity shear kelvin--helmholtz instabilities in strongly rotating
  tokamak plasmas,  \emph{Nucl. Fusion} \textbf{52} no.~4 (2012), 042005.
  \doi{10.1088/0029-5515/52/4/042005}.

\bibitem[DG66]{dangelo66}
\bgroup\scshape{}N.~D'angelo\egroup{} and \bgroup\scshape{}S.~V.
  Goeler\egroup{}, Investigation of the kelvin-helmholtz instability in a
  cesium plasma,  \emph{Phys. Fluids} \textbf{9} no.~2 (1966), 309--313.
  \doi{10.1063/1.1761674}.

\bibitem[D'A65]{dangelo65}
\bgroup\scshape{}N.~D'Angelo\egroup{}, Kelvin—helmholtz instability in a
  fully ionized plasma in a magnetic field,  \emph{Phys. Fluids} \textbf{8}
  no.~9 (1965), 1748--1750. \doi{10.1063/1.1761496}.

\bibitem[DHG{\etalchar{+}}92]{drake92}
\bgroup\scshape{}J.~Drake\egroup{}, \bgroup\scshape{}A.~Hassam\egroup{},
  \bgroup\scshape{}P.~Guzdar\egroup{}, \bgroup\scshape{}C.~Liu\egroup{}, and
  \bgroup\scshape{}D.~McCarthy\egroup{}, Loss of static equilibrium, flow
  generation and the development of turbulence at the edge of tokamaks,
  \emph{Nucl. Fusion} \textbf{32} no.~9 (1992), 1657.
  \doi{10.1088/0029-5515/32/9/I15}.

\bibitem[FDG{\etalchar{+}}12]{field12}
\bgroup\scshape{}A.~Field\egroup{}, \bgroup\scshape{}D.~Dunai\egroup{},
  \bgroup\scshape{}R.~Gaffka\egroup{}, \bgroup\scshape{}Y.-c. Ghim\egroup{},
  \bgroup\scshape{}I.~Kiss\egroup{}, \bgroup\scshape{}B.~Meszaros\egroup{},
  \bgroup\scshape{}T.~Krizsanoczi\egroup{},
  \bgroup\scshape{}S.~Shibaev\egroup{}, and
  \bgroup\scshape{}S.~Zoletnik\egroup{}, Beam emission spectroscopy turbulence
  imaging system for the mast spherical tokamak,  \emph{Rev. Sci. Instr.}
  \textbf{83} no.~1 (2012), 013508. \doi{10.1063/1.3669756}.

\bibitem[GSG{\etalchar{+}}02]{garbet02}
\bgroup\scshape{}X.~Garbet\egroup{}, \bgroup\scshape{}Y.~Sarazin\egroup{},
  \bgroup\scshape{}P.~Ghendrih\egroup{}, \bgroup\scshape{}S.~Benkadda\egroup{},
  \bgroup\scshape{}P.~Beyer\egroup{}, \bgroup\scshape{}C.~Figarella\egroup{},
  and \bgroup\scshape{}I.~Voitsekhovitch\egroup{}, Turbulence simulations of
  transport barriers with toroidal velocity,  \emph{Phys. Plasmas} \textbf{9}
  no.~9 (2002), 3893--3905. \doi{10.1063/1.1499494}.

\bibitem[GCH{\etalchar{+}}19]{grierson19}
\bgroup\scshape{}B.~Grierson\egroup{}, \bgroup\scshape{}C.~Chrystal\egroup{},
  \bgroup\scshape{}S.~Haskey\egroup{}, \bgroup\scshape{}W.~Wang\egroup{},
  \bgroup\scshape{}T.~Rhodes\egroup{}, \bgroup\scshape{}G.~McKee\egroup{},
  \bgroup\scshape{}K.~Barada\egroup{}, \bgroup\scshape{}X.~Yuan\egroup{},
  \bgroup\scshape{}M.~Nave\egroup{}, \bgroup\scshape{}A.~Ashourvan\egroup{},
  and \bgroup\scshape{}C.~Holland\egroup{}, Main-ion intrinsic toroidal
  rotation across the itg/tem boundary in diii-d discharges during ohmic and
  electron cyclotron heating,  \emph{Phys. Plasmas} \textbf{26} no.~4 (2019),
  042304. \doi{10.1063/1.5090505}.

\bibitem[GWZ19]{guo19}
\bgroup\scshape{}W.~Guo\egroup{}, \bgroup\scshape{}L.~Wang\egroup{}, and
  \bgroup\scshape{}G.~Zhuang\egroup{}, Impurity transport driven by parallel
  velocity shear turbulence in hydrogen isotope plasmas,  \emph{Nucl. Fusion}
  \textbf{59} no.~7 (2019), 076012. \doi{10.1088/1741-4326/ab1967}.

\bibitem[HSC{\etalchar{+}}12]{highcock12}
\bgroup\scshape{}E.~Highcock\egroup{},
  \bgroup\scshape{}A.~Schekochihin\egroup{},
  \bgroup\scshape{}S.~Cowley\egroup{}, \bgroup\scshape{}M.~Barnes\egroup{},
  \bgroup\scshape{}F.~Parra\egroup{}, \bgroup\scshape{}C.~Roach\egroup{}, and
  \bgroup\scshape{}W.~Dorland\egroup{}, Zero-turbulence manifold in a toroidal
  plasma,  \emph{Phys. Rev. Lett.} \textbf{109} no.~26 (2012), 265001.
  \doi{10.1103/PhysRevLett.109.265001}.

\bibitem[IKK{\etalchar{+}}16]{inagaki16}
\bgroup\scshape{}S.~Inagaki\egroup{}, \bgroup\scshape{}T.~Kobayashi\egroup{},
  \bgroup\scshape{}Y.~Kosuga\egroup{}, \bgroup\scshape{}S.-I. Itoh\egroup{},
  \bgroup\scshape{}T.~Mitsuzono\egroup{},
  \bgroup\scshape{}Y.~Nagashima\egroup{}, \bgroup\scshape{}H.~Arakawa\egroup{},
  \bgroup\scshape{}T.~Yamada\egroup{}, \bgroup\scshape{}Y.~Miwa\egroup{},
  \bgroup\scshape{}N.~Kasuya\egroup{}, and \bgroup\scshape{}others\egroup{}, A
  concept of cross-ferroic plasma turbulence,  \emph{Sci. Rep.} \textbf{6}
  no.~1 (2016), 22189. \doi{10.1038/srep22189}.

\bibitem[KTH03]{kaneko03}
\bgroup\scshape{}T.~Kaneko\egroup{}, \bgroup\scshape{}H.~Tsunoyama\egroup{},
  and \bgroup\scshape{}R.~Hatakeyama\egroup{}, Drift-wave instability excited
  by field-aligned ion flow velocity shear in the absence of electron current,
  \emph{Phys. Rev. Lett.} \textbf{90} no.~12 (2003), 125001.
  \doi{10.1103/PhysRevLett.90.125001}.

\bibitem[KIK{\etalchar{+}}16]{kobayashi16}
\bgroup\scshape{}T.~Kobayashi\egroup{}, \bgroup\scshape{}S.~Inagaki\egroup{},
  \bgroup\scshape{}Y.~Kosuga\egroup{}, \bgroup\scshape{}M.~Sasaki\egroup{},
  \bgroup\scshape{}Y.~Nagashima\egroup{}, \bgroup\scshape{}T.~Yamada\egroup{},
  \bgroup\scshape{}H.~Arakawa\egroup{}, \bgroup\scshape{}N.~Kasuya\egroup{},
  \bgroup\scshape{}A.~Fujisawa\egroup{}, \bgroup\scshape{}S.-I. Itoh\egroup{},
  and \bgroup\scshape{}others\egroup{}, Structure formation in parallel ion
  flow and density profiles by cross-ferroic turbulent transport in linear
  magnetized plasma,  \emph{Phys. Plasmas} \textbf{23} no.~10 (2016), 102311.
  \doi{10.1063/1.4965915}.

\bibitem[KBLO24]{kosuga24}
\bgroup\scshape{}Y.~Kosuga\egroup{}, \bgroup\scshape{}J.~Bourgeois\egroup{},
  \bgroup\scshape{}M.~Lesur\egroup{}, and \bgroup\scshape{}I.~Oyama\egroup{},
  Breathing impure plasmas,  \emph{Plasma Phys. Control. Fusion} \textbf{66}
  no.~7 (2024), 075018. \doi{10.1088/1361-6587/ad5105}.

\bibitem[KII17]{kosuga17pop}
\bgroup\scshape{}Y.~Kosuga\egroup{}, \bgroup\scshape{}S.-I. Itoh\egroup{}, and
  \bgroup\scshape{}K.~Itoh\egroup{}, Zonal flow generation in parallel flow
  shear driven turbulence,  \emph{Phys. Plasmas} \textbf{24} no.~3 (2017),
  032304. \doi{10.1063/1.4978485}.

\bibitem[KII15]{kosuga15}
\bgroup\scshape{}Y.~Kosuga\egroup{}, \bgroup\scshape{}S.-I. Itoh\egroup{}, and
  \bgroup\scshape{}K.~Itoh\egroup{}, Density peaking by parallel flow shear
  driven instability,  \emph{Plasma Fusion Research} \textbf{10} (2015),
  3401024--3401024. \doi{10.1585/pfr.10.3401024}.

\bibitem[LBK23]{lesur23iaea}
\bgroup\scshape{}M.~Lesur\egroup{}, \bgroup\scshape{}J.~Bourgeois\egroup{}, and
  \bgroup\scshape{}Y.~Kosuga\egroup{}, Impurity parallel velocity gradient
  instability,  in \emph{Fusion Energy 2023 (Proc.~29th Int.~Conf.~London)},
  Vienna: IAEA, International Centre, 2023.

\bibitem[McC02]{mccarthy02}
\bgroup\scshape{}D.~McCarthy\egroup{}, Edge harmonic oscillations produced by
  toroidal velocity shear,  \emph{Phys. Plasmas} \textbf{9} no.~6 (2002),
  2451--2454. \doi{10.1063/1.1472503}.

\bibitem[NCL10]{newton10}
\bgroup\scshape{}S.~L. Newton\egroup{}, \bgroup\scshape{}S.~C. Cowley\egroup{},
  and \bgroup\scshape{}N.~F. Loureiro\egroup{}, Understanding the effect of
  sheared flow on microinstabilities,  \emph{Plasma Phys. Control. Fusion}
  \textbf{52} no.~12 (2010), 125001. \doi{10.1088/0741-3335/52/12/125001}.

\bibitem[SHC12]{schekochihin12}
\bgroup\scshape{}A.~Schekochihin\egroup{},
  \bgroup\scshape{}E.~Highcock\egroup{}, and
  \bgroup\scshape{}S.~Cowley\egroup{}, Subcritical fluctuations and suppression
  of turbulence in differentially rotating gyrokinetic plasmas,  \emph{Plasma
  Phys. Control. Fusion} \textbf{54} no.~5 (2012), 055011.
  \doi{10.1088/0741-3335/54/5/055011}.

\bibitem[TGF{\etalchar{+}}02]{testa02}
\bgroup\scshape{}D.~Testa\egroup{}, \bgroup\scshape{}C.~Giroud\egroup{},
  \bgroup\scshape{}A.~Fasoli\egroup{}, \bgroup\scshape{}K.-D. Zastrow\egroup{},
  and \bgroup\scshape{}E.-J. Team\egroup{}, On the measurement of toroidal
  rotation for the impurity and the main ion species on the joint european
  torus,  \emph{Phys. Plasmas} \textbf{9} no.~1 (2002), 243--250.
  \doi{10.1063/1.1427727}.

\bibitem[WWY{\etalchar{+}}98]{wang98}
\bgroup\scshape{}G.~Wang\egroup{}, \bgroup\scshape{}L.~Wang\egroup{},
  \bgroup\scshape{}X.~Yang\egroup{}, \bgroup\scshape{}C.~Feng\egroup{},
  \bgroup\scshape{}D.~Jiang\egroup{}, and \bgroup\scshape{}X.~Qi\egroup{},
  Evidence for parallel flow shear instability in the edge plasma of the ct-6b
  tokamak,  \emph{Plasma Phys. Control. Fusion} \textbf{40} no.~3 (1998), 429.
  \doi{10.1088/0741-3335/40/3/007}.

\bibitem[WER{\etalchar{+}}15]{wang15}
\bgroup\scshape{}W.~Wang\egroup{}, \bgroup\scshape{}S.~Ethier\egroup{},
  \bgroup\scshape{}Y.~Ren\egroup{}, \bgroup\scshape{}S.~Kaye\egroup{},
  \bgroup\scshape{}J.~Chen\egroup{}, \bgroup\scshape{}E.~Startsev\egroup{}, and
  \bgroup\scshape{}Z.~Lu\egroup{}, Distinct turbulence sources and confinement
  features in the spherical tokamak plasma regime,  \emph{Nucl. Fusion}
  \textbf{55} no.~12 (2015), 122001. \doi{10.1088/0029-5515/55/12/122001}.

\end{thebibliography}
\addcontentsline{toc}{section}{References}

\end{document}